\documentclass[aip,amsmath,amssymb,reprint]{revtex4-2}
\usepackage[utf8]{inputenc}
\usepackage[english]{babel}

\usepackage{epsfig}
\usepackage{wrapfig}
\usepackage{bbm}
\usepackage[usenames]{color}
\usepackage{array}
\usepackage{times}
\usepackage{bm}
\usepackage[normalem]{ulem}
\usepackage{graphicx}
\usepackage{physics}
\usepackage{svg}
\usepackage{float}
\usepackage{multirow}
\usepackage[breaklinks=true]{hyperref}
\usepackage{xcolor}
\usepackage{amsmath,amsfonts,amssymb}
\usepackage{graphicx}
\usepackage{hyperref}
\usepackage{color}
\usepackage{stackengine}
\usepackage{subfigure}
\usepackage{verbatim}
\usepackage{comment}
\usepackage{stmaryrd} 


\begin{document}


\title{Quantized topological transport mediated by the long-range couplings}

\author{Ekaterina S. Lebedeva}
\affiliation{School of Physics and Engineering, ITMO University, Saint-Petersburg 197101, Russia}

\author{Maxim Mazanov}%
\affiliation{School of Physics and Engineering, ITMO University, Saint-Petersburg 197101, Russia}

\author{Alexey V. Kavokin}
\affiliation{Abrikosov Center for Theoretical Physics, Moscow Institute of Physics and Technology, Dolgoprudny, Moscow Region 141701, Russia}
\affiliation{Russian Quantum Center, 30-1, Bolshoy boulevard, Skolkovo, Moscow Region, Russia}
\affiliation{School of Science, Westlake University, 18 Shilongshan Road, Hangzhou 310024, Zhejiang Province, China}

\author{Maxim A. Gorlach}%
 \email{m.gorlach@metalab.ifmo.ru}
\affiliation{School of Physics and Engineering, ITMO University, Saint-Petersburg 197101, Russia}

\date{\today}

\begin{abstract}
Certain topological systems with time-varying Hamiltonian enable quantized and disorder-robust transport of excitations. Here, we introduce the modification of the celebrated Thouless pump when the on-site energies remain fixed, while the nearest and next-nearest neighbor couplings vary in time. We demonstrate quantized transport of excitations and propose an experimental implementation using an array of evanescently coupled optical waveguides. 
\end{abstract}

\maketitle

Topological physics uncovers promising approaches to control the localization and propagation of light by tailoring the bandstructure of the material and harnessing localized or propagating topological modes~\cite{Lu2014,Ozawa2019,Price2022}. 

Even richer physics arises when the Hamiltonian of the system varies in time. If such variation is periodic, this gives rise to Floquet physics and non-equilibrium phases with tailored properties~\cite{Goldman2014,Rechtsman2013}.  Of special interest is a periodic variation of the Hamiltonian resulting in the transport of excitations analogously to the water flow driven by the Archimedean screw. Leveraging the topological nature of the system such transport can be made disorder-resilient in the sense that the charge transferred during a single pumping cycle is robustly quantized~\cite{BookShen,Asboth}.

Historically the first example of such topological transport was the so-called Thouless pump~\cite{Thouless1983} which utilized the Rice-Mele model~\cite{BookShen,Asboth} with time-varying on-site energies and couplings between the sites. The latter system, in turn, is a generalization of the celebrated Su-Schrieffer-Heeger model~\cite{Su1979Jun}, a paradigmatic example of a one-dimensional topological system.

The Thouless pump has been realized experimentally for ultracold atoms in a dynamically controlled optical lattice~\cite{Lohse2015,Nakajima2016,Nakajima2021} as well as for photonic systems~\cite{Kraus2012,Tang2016,Fedorova2020,Cerjan2020}. Recent generalizations include systems with strong nonlinearity~\cite{Sebabrata2021,Fu2022,Fu2022a,Sebabrata2023,Ravets2025}, fractional Thouless pumping~\cite{Tao2025a}, pumping of two-dimensional~\cite{Wang2022} and multiband systems~\cite{You2022,Sun2024,Tao2025} which could feature non-Abelian physics~\cite{You2022,Sun2024}. Importantly, the protocol of Thouless pumping in simple two-band systems requires a synchronized variation of both couplings and on-site energies, which is quite challenging to implement in the arrays of evanescently coupled optical waveguides requiring a simultaneous modulation of the refractive index contrast and distance between the waveguides~\cite{Cerjan2020}.

In this Letter we address that challenge and design an alternative protocol of topological pumping which only requires time-varying couplings. To compensate for the lack of time-varying on-site energies, we introduce next-nearest neighbor couplings. Below, we analyze this pumping scheme and demonstrate that it can be readily implemented using the arrays of laser-written evanescently coupled optical waveguides without the need to modulate their refractive index along the direction of propagation. In addition, we outline a prospective realization of this physics for another material platform~-- polariton condensates.

\begin{figure}[ht!]
    \centering
    \includegraphics[width=0.65\linewidth]{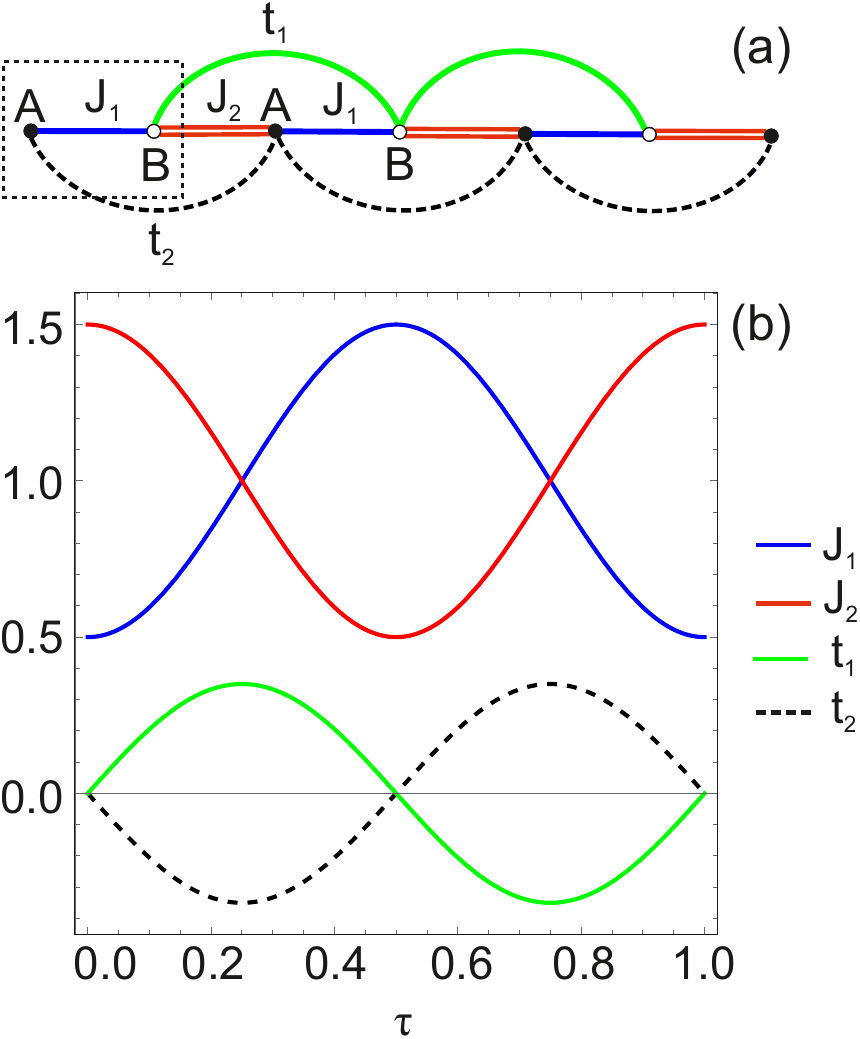}
    \caption{(a) Schematic of the studied 1D lattice. Synchronized variation of nearest and next-nearest-neighbor couplings enables quantized topological transport. Dashed rectangle shows the unit cell including the sites of A and B sublattices. (b) Evolution of the couplings during one pumping cycle. Parameters: 
    $J_0 = 1$ , $A = 1$, $B=0.35$. 
    }
    \label{Fig.1}
\end{figure}

Specifically, we consider a one-dimensional lattice with the nearest-neighbor couplings $J_1$ and $J_2$ resembling the celebrated Su-Schrieffer-Heeger 
model. In addition, the structure is supplemented by the next-nearest neighbor couplings $t_1$ and $t_2$ connecting the sites of the same sublattice [Fig.~\ref{Fig.1}(a)]. Hence, Bloch Hamiltonian of the periodic system is presented in the form
\begin{equation}
\label{Ham1}
H(k)=\begin{pmatrix}
2 t_2 \cos k & J_1+J_2 e^{-i k} \\
J_1+J_2 e^{i k} & 2 t_1 \cos k
\end{pmatrix}\:.
\end{equation}
To realize pumping, the Hamiltonian varies in time  $\tau = t/T$,  $T$ being the period of the pumping cycle, as follows:
\begin{align}
J_1(\tau) &= J_0 - \frac{A}{2}\cos(2\pi \tau)\:, \label{eq:J1}\\
J_2(\tau) &= J_0 + \frac{A}{2}\cos(2\pi \tau)\:, \label{eq:J2}\\
t_1(\tau) &= B \sin(2\pi \tau)\:, \label{eq:t1}\\
t_2(\tau) &= -B \sin(2\pi \tau)\:,\label{eq:t2}
\end{align}
%
where $J_0$, $A$, and $B$ are constant factors defining the modulation amplitude. Figure~\ref{Fig.1}(b) illustrates the dependence of the couplings $J_{1,2}$, $t_{1,2}$ on time. While all couplings vary harmonically, there is a $\pi/2$ phase shift between the nearest and next-nearest neighbor couplings. In addition, nearest-neighbor couplings $J_{1,2}$ remain positive throughout the entire pumping cycle, while the next-nearest neighbor couplings $t_{1,2}$ switch their sign.

Inspecting Bloch Hamiltonian Eq.~\eqref{Ham1}, we observe that it is quite similar to that in the standard Thouless pump protocol. A formal difference appears in $\cos k$ term at the diagonal. However, this leads to a quite different physical realization which we investigate below.

While the time-varying Hamiltonian has no stationary states, it is instructive to diagonalize it at the arbitrary moment of time $\tau$ and evaluate its instantaneous spectrum given by the expression
%
%
\begin{equation}\label{eq:spectrum}
\begin{split}
& E(k)=(t_1+t_2)\cos{k} \\
& \pm \sqrt{J_1^2+J_2^2+2J_1J_2\cos{k}+(t_1-t_2)\cos^2{k}},
\end{split}
\end{equation}
which features two bulk bands separated by the bandgap, while the couplings $J_{1,2}$ and $t_{1,2}$ depend on time according to Eqs.~\eqref{eq:J1}-\eqref{eq:t2}. 

\begin{figure}[ht!] 
    \centering
    \includegraphics[width=0.7\linewidth]{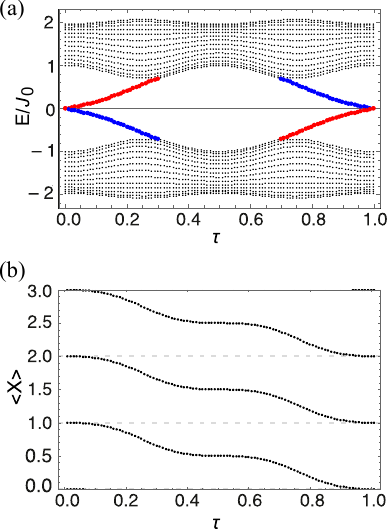}
    \caption{Instantaneous spectrum and Wannier centers of the driven system. (a) Evolution of the instantaneous spectrum calculated for a finite system consisting of $N=14$ unit cells with open boundary conditions. 
    Left- and right-localized edge states are highlighted by red and blue, respectively. 
    (b) Evolution of the instantaneous Wannier centers for the same system with periodic boundary conditions} during a single driving cycle. 
    
    \label{Fig.2}
\end{figure}

The spectrum for a finite 14-unit-cell lattice with the open boundary conditions in Fig.~\ref{Fig.2}(a) provides an intuition on how the system behaves in the adiabatic limit, i.e. when the driving frequency is much smaller than the characteristic eigenfrequencies of the system. 
%
%
The results suggest that the gap in the spectrum remains open throughout the entire pumping cycle, and there are only two edge states which cross the bandgap and exhibit nonzero group velocity. As we discuss in the Supplementary Materials, these edge states are a signature of the topological tranport.


A complementary perspective is obtained by studying the same lattice but with the periodic boundary conditions.
While Bloch functions do not provide immediate insights into the properties of the pumping scheme, their linear combinations known as Wannier functions can be readily visualized; they feature good localization, form an orthogonal set and are related to each other via translation~\cite{Asboth}.

To construct the Wannier functions, we first introduce the position operator in the form suitable for the periodic systems~\cite{Resta1998Mar,Marzari1997Nov,Marzari2012Oct}
\begin{equation}
    \hat{X}=e^{\frac{2 \pi i}{N}\hat{x}},
\end{equation}
where $\hat{x}$ yields the number of the unit cell 
$1,2,\dots N$. The operator $\hat{X}$ defined in this way respects the periodic boundary conditions, while the expectation values of the particle position are computed from its eigenvalues $X_m$ as $-iN/(2\pi)\,\log(X_m)$.

Next we project this position operator onto the lowest  band using the projector
\begin{equation}
 \hat{P} = \sum_{n=1}^{N/2} | \Psi_n \rangle \langle \Psi_n |\:,
\end{equation}
$\ket{\Psi_n}$ being $n^{\text{th}}$ eigenstate from the respective band:
\begin{equation}
    \hat{X}_p=\hat{P}\hat{X}\hat{P}\:.
\end{equation}

The eigenvectors of the projected position operator $\hat{X}_p$ provide the Wannier functions of the lowest band~\cite{Asboth}. At the same time, the eigenvalues of $\hat{X}_p$, $\lambda_m$, define so-called Wannier centers via
\begin{equation}
    W_m=\frac{N}{2 \pi i}\log{\lambda_m}\:.
\end{equation}
We plot the instantaneous Wannier centers for our system in Fig.~\ref{Fig.2}(b) and observe that all Wannier centers shift by one unit cell during the pumping cycle. Since an arbitrary input state from the lowest band can be decomposed into the superposition of Wannier functions, this suggests that the designed protocol transports any such state by one unit cell during a single pumping cycle. The topological nature of this pumping scheme can be further highlighted by computing the space-time Chern number of the lowest band which is $C=1$ (see Supplementary Materials).


\begin{figure}[ht!]
    \centering
    \includegraphics[width=0.65\linewidth]{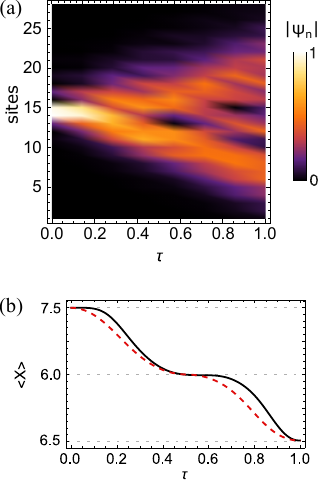}
    \caption{(a) Diffraction of the point-like wavepacket projected onto the lower Bloch band for the 14 unit cell lattice. (b) Displacement of the wavepacket center of mass by $1.005$ unit cell over one cycle $ T = 7 $ (black) and corresponding Wannier center trajectory (red dashed).}
    \label{Fig.3}
\end{figure}


So far we discussed the adiabatic limit of our model. To assess how the transport protocol works at a finite driving frequency, we solve the temporal evolution of the system consisting of $N=14$ unit cells with the open boundary conditions. We choose the driving period $T = 7$, which substantially exceeds the average inverse bandgap size $ 1/\delta \simeq 0.6 $, justifying the adiabatic approximation. 
We assume that the initial state is localized in the middle of the array and belongs entirely to the lowest band, i.e. 
\begin{equation}
    \ket{\psi(0)}=\hat{P}\,\left(0,\dots, 1,\dots, 0\right)^T\:.
\end{equation}
%

Solving the Schr\"odinger equation, we recover the probability distribution in Fig.~\ref{Fig.3}(a) which shows that initially localized wavepacket starts to spread over the lattice. While the distribution $|\psi_n(t)|^2$ itself does not exhibit clear signatures of transport, it is instructive to track the motion of the wavepacket center of mass, i.e. $\left<x(t)\right>=\left<\psi\left|\hat{x}\right|\psi\right>$ [Fig.~\ref{Fig.3}(b)]. We observe that the center-of-mass motion resembles the temporal dependence of the Wannier centers and exhibits a shift equal to 1.005 unit cell during a single pumping cycle. Nearly quantized change of $\left<x\right>$ is due to the topological nature of the pump, while slight violation of the quantization appears due to the non-adiabatic evolution of the system. Note that in order to have strictly quantized pumping at a finite driving frequency the temporal profile of the driving has to be fine-tuned~\cite{Malikis2022} using, for instance, counter-adiabatic driving method~\cite{Angelis2020,Ke2025}. Due to the chosen timespan, the wavepacket initialized in the middle of the lattice practically does not interact with the edges and thus open boundary conditions do not destroy quantized transport.




\begin{figure*}[ht!]
    \centering
    \includegraphics[width=0.99\linewidth]{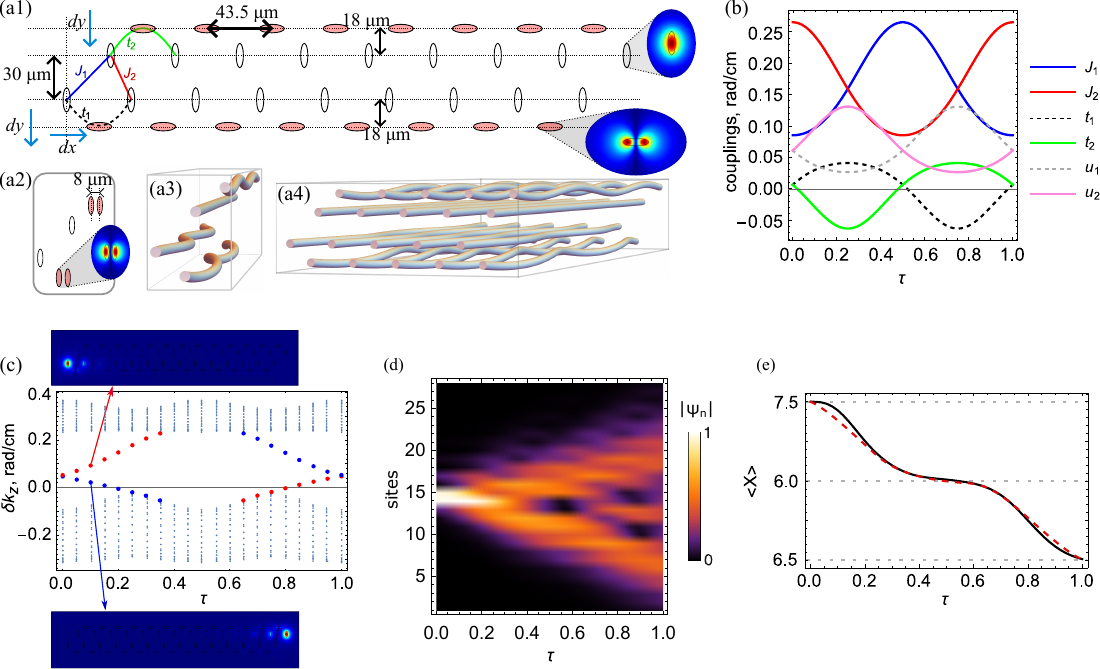}
    \caption{
    (a) Chosen geometry of the lattice and its spatial modulation. Insets show electric field profiles (amplitude) of symmetric and antisymmetric modes in vertical and detuned horizontal waveguides, respectively. 
    Insets illustrate the choice of the unit cell with identical spacings and with $p$-mode photonic molecule connectors~(a2), 
    the adiabatic modulation of waveguides along their propagation coordinate $z$ for one unit cell~(a3) and a $5$-unit-cell lattice~(a4). 
    (b) Modulation of the couplings $J_{1,2}$, $t_{1,2}$ and detunings $u_{1,2}$ extracted from numerically simulated splitting of the dimer eigenmodes. 
    (c) Instantaneous spectrum of the propagation constants computed for a finite lattice with $14$ unit cells. Left- and right-localized edge states are highlighted by red and blue, respectively. 
    (d) Tight-binding simulation of the real-space discrete diffraction pattern for the initial point-like excitation projected onto the lowest Bloch band of a $14$ unit cell lattice and modulation period $L = 50$~cm. 
    (e) Time dependence of the center of mass of the intensity distribution (black solid line), displaced by $1.0068$ unit cells over one modulation period. The trajectory of the Wannier center is shown by the dashed red curve for comparison.
    }
    \label{Fig.4}
\end{figure*}

As a specific platform to implement our protocol of topological pumping at optical wavelengths, we propose an optical waveguide lattice fabricated via  femtosecond laser writing technique~\cite{Szameit2010Jul,Wang2021Mar,Yan2024Oct}. 
We exploit a formal analogy between the tight-binding Schr\"odinger equation and the coupled-mode description of the waveguide lattice: the time variable $t = \tau T$ in the former corresponds to the propagation distance $z = \tau L$ in the waveguide system, where $L$ is the spatial period of the modulation along the waveguides.

We propose the zigzag-like geometry shown in Fig.~\ref{Fig.4}(a) consisting of the main vertical waveguides shown in white and detuned connector waveguides depicted in red. The positioning and orientation of connector waveguides is chosen in such a way that they efficiently couple to the sublattice of main waveguides close to them, but remain practically decoupled from another sublattice of main waveguides. 
Depending on parameters, the waveguides can support the modes with the different symmetry of the near field profile. To enable larger bandgap, we require that the connector waveguides support dipolar $p_x$ modes spectrally close to the $s$ modes of the main waveguides (see Supplementary Materials).
%
%
This realizes an instance of multi-orbital physics~\cite{Vicencio2025Jul,Guzman-Silva2021Aug,Caceres-Aravena2022Jun,Caceres-Aravena2024Sep,Mazanov2024Apr,Caceres-Aravena2019Jul,Caceres-Aravena2020Aug,Savelev2020Oct,Mikhin2023Mar,Gorbach2023Apr,Liu2024Dec,Gao2024Mar,Rajeevan2025Jan,Mazanov2024May,Li2016Sep,Pelegri2019Feb}. 
As the fabrication of elliptical waveguides with the different ellipse orientations is challenging, for the connector $p$-mode sites one could use fine-tuned photonic molecules based on pairs of closely placed vertically oriented waveguides~\cite{Mazanov2024Apr}.



%

Since the connector waveguides are assumed to be detuned from the main ones, they realize non-resonant coupling and can be excluded from the description using the degenerate perturbation theory~\cite{Bir1974,Gorlach2018Mar,Mittal2019Oct,Mukherjee2018Aug}~We illustrate this procedure for a trimer of two $s$-mode main waveguides connected by the detuned $p$-mode waveguide, described by the Hamiltonian 
\begin{equation}
\label{HamTrimerpMAIN}
H_{\text{trimer}}
=
\begin{pmatrix}
0 & \kappa & \gamma \\
\kappa & 0 & -\gamma \\
\gamma & -\gamma & \Delta 
\end{pmatrix},
\end{equation}
where $\gamma$ is the absolute value of the coupling between main and detuned waveguides, $\kappa$ is the direct coupling between the two main waveguides, and $\Delta = k_{z}^{p} - k_{z}^{s}$ is the propagation constants detuning. 
Assuming $|\Delta| \gg \kappa, \gamma$, we exclude the detuned $p$-mode~\cite{Bir1974,Gorlach2018Mar} and 
obtain the effective $2 \times 2$ 
effective Hamiltonian 
\begin{equation}
H_{\text{dimer}}
=
\begin{pmatrix}
u & t \\
t & u 
\end{pmatrix}
\end{equation}
with effective 
next-nearest-neighbor couplings $t = \kappa + \gamma^2 / (\kappa + \Delta)$ 
and detunings $u = - \gamma^2/(\kappa + \Delta)$, as further discussed in the Supplementary Materials.

%
Importantly, the antisymmetric shape of $p$-like modes ensures that the couplings $t_{1,2}$ and detunings $u_{1,2}$ for each sublattice are modulated out of phase. As a result, the two mechanisms of bandgap opening add up, leading to the larger bandgap and allowing to reduce the modulation period.

In our simulations, we choose the wavelength $\lambda = 730$~nm and ambient glass refractive index $n = 1.48$ (borosilicate) with elliptical waveguide profiles with a base contrast of $\delta n = 4 \cdot 10^{-4}$ and semi-axes $a=2.45\,\mu$m and $b=8.18\,\mu$m~\cite{Guzman-Silva2021Aug,Caceres-Aravena2022Jun}. 
%
We choose detuning of the connector $p$-mode waveguides $\Delta = -3$~rad/cm by setting their refractive index contrast to $\delta n = 9.0 \cdot 10^{-4}$, which translates into the increase in the laser writing power by $125 \%$ during the fabrication of connector waveguides with respect to the main lattice waveguides. 
As we assume $p$-mode waveguides have geometrically identical but $90^\circ$-rotated elliptic profiles compared to main $s$-mode ones, a possible fabrication technique could utilize 
$p$-mode photonic molecule connectors~\cite{Mazanov2024Apr} comprising pairs of same waveguides separated by $8\,\mu$m with a contrast $\delta n = 5.73 \cdot 10^{-4}$, written from a single glass wafer facet in the same lattice geometry, see~Fig.~\ref{Fig.4}(a2). This produces almost identical effective couplings and detunings $\gamma, u$. 
%

To create the modulation of the couplings, we assume that the waveguides are adiabatically curved along their propagation direction~\cite{Rechtsman2013} to create harmonic modulations of the distances $dx$ and $dy$ between the adjacent waveguides 
%
Specifically, the coordinates of the lower main lattice waveguides and lower connector waveguides are modulated horizontally as $dx = - 8 \cos{(2 \pi \tau)} \,\,\, \text{$\mu$m}$, while the coordinates of both lower and upper connector waveguides are additionally modulated vertically as $dy = - 4 \sin{(2 \pi \tau)} \,\,\, \text{$\mu$m}$. 

Such bending of the waveguides gives rise to the $z$-dependent couplings 
as further analyzed in the Supplementary Materials. To calculate these dependencies, we first obtain the couplings $J (dx)$, $\gamma (dy)$, $\kappa$ at perfect degeneracy ($\Delta = 0$) as functions of respective waveguide spatial displacements 
from the eigenmodes of respective waveguide pairs, tracking the splitting between the modes which quantifies the strength of the couplings. 
Substituting $\gamma (dy)$, $\kappa$ and $\Delta$ into the effective couplings $t_{1,2}$ and detunings $u_{1,2}$ and using specific modulations $dx(\tau), dy(\tau)$, we arrive to the results in Fig.~\ref{Fig.4}(b).

Next we simulate numerically the spectrum for a lattice consisting of $14$ unit cells [Fig.~\ref{Fig.4}(c)] and observe a good agreement with the tight-binding result using numerically calculated couplings (see Supplementary Materials for details). 
The spectrum shows left-and right-localized topological edge states traversing the complete bandgap in the opposite directions, indicating the presence of topological pump.  
Moreover, the spectrum in this case features a particularly large bandgap, facilitating quantized transport for smaller lattices and shorter modulation periods with the greater robustness to the disorder.
%

Finally, the tight-binding simulation of topological pumping in Fig.~\ref{Fig.4}(d) shows the asymmetric spreading of the initial point-like excitation projected onto the lower Bloch band 
for the array consisting of $14$ unit cells with the modulation period $L = 50$~cm. 
The center-of-mass trajectory of the intensity distribution in Fig.~\ref{Fig.4}(e) (black curve) shows quantized transport and follows quite closely the trajectory of the respective Wannier center (red dashed curve) with the total shift of $1.0068$ unit cells per modulation period. 
Figures~\ref{Fig.4}(d,e) feature many parallels with Fig.~\ref{Fig.3}(a), even though the former describes a realistic model with the detunings and non-harmonic coupling dependencies on $\tau$, while the latter~-- a simple model with harmonic coupling modulation without detunings. This can be attributed to the adiabatic regime of the pumping with similar adiabaticity parameters $\alpha = \delta L$, where $\delta$ is the bandgap size, as well as the validity of the degenerate perturbation theory.


For smaller modulation periods, quantized nature of the transport is violated resulting, for instance, in the shift of $0.84$ unit cell for the modulation period  $L=20$~cm. 
Note that the lattice periods required for quantized adiabatic transport may be further decreased for lattices operating close to the flat-band or all-bands-flat conditions~\cite{Caceres-Aravena2022Jun}, and also in the nonlinear pumping regime when the diffraction of the wave packet is suppressed by the nonlinearity~\cite{Sebabrata2021}.


In summary, we have proposed a topological pump allowing to transfer the quantum state by varying nearest-neighbor and next-nearest-neighbor couplings. This protocol provides an alternative to the celebrated Thouless pump and does not involve the change of on-site energies, which is especially suitable for the arrays of evanescently coupled optical waveguides. The proposed scheme is not limited to optical waveguides and 
can be readily generalized to the other physical platforms, e.g. to the polariton condensates trapped in a slowly varying periodic potential~\cite{Gnusov2023,Fraser2024} with a conceptual scheme elaborated in the Supplementary Materials.

See Supplementary Materials for the details of numerical simulations of the system with symmetric and anti-symmetric connector modes in an optical waveguide lattice, including lattice geometry and chosen modulation scheme, application of the degenerate perturbation theory for connector sites, evaluation of the effective couplings and detuning of connector waveguides, comparison of the tight-binding and numerical finite-lattice spectra and tight-binding simulations of quantized pumping, proposal for photonic molecule connector waveguides, results for the full tight-binding model, as well the proposed implementation for the lattice of coupled polariton condensates.


We acknowledge Andrei Stepanenko and Anton Nalitov for valuable discussions. This work was supported by the Russian Science Foundation grant No.~25-79-31027.


\section*{Author declarations}
{\bf Conflict of interest}

The authors have no conflicts to disclose.

\section*{Data Availability Statement}
The data that support the findings of this study are available from the corresponding author upon reasonable request.


\bibliography{QuantizedTransport}

\end{document}


\title{Supplemental Materials: \\ Quantized topological transport mediated by the long-range couplings
}

\author{Ekaterina S. Lebedeva}

\affiliation{School of Physics and Engineering, ITMO University, Saint Petersburg 197101, Russia}

\author{Maxim Mazanov}
\affiliation{School of Physics and Engineering, ITMO University, Saint Petersburg 197101, Russia}

\author{Alexey V. Kavokin}
\affiliation{Abrikosov Center for Theoretical Physics, Moscow Institute of Physics and Technology, Dolgoprudny, Moscow Region 141701, Russia}
\affiliation{Russian Quantum Center, 30-1, Bolshoy boulevard, Skolkovo, Moscow Region, Russia}
\affiliation{School of Science, Westlake University, 18 Shilongshan Road, Hangzhou 310024, Zhejiang Province, China}

\author{Maxim A. Gorlach}
\affiliation{School of Physics and Engineering, ITMO University, Saint Petersburg 197101, Russia}
\email{m.gorlach@metalab.ifmo.ru}

\maketitle

\onecolumngrid

\setcounter{equation}{0}
\setcounter{figure}{0}
\setcounter{table}{0}
\setcounter{page}{1}
\setcounter{section}{0}
\makeatletter
\renewcommand{\theequation}{S\arabic{equation}}
\renewcommand{\thefigure}{S\arabic{figure}}
\renewcommand{\bibnumfmt}[1]{[S#1]}
\renewcommand{\citenumfont}[1]{S#1}

\tableofcontents


\section{I. Geometry of the waveguide lattice and its spatial modulations. Degenerate perturbation theory for connector sites} 

\begin{figure}[h!]
    \centering
    \includegraphics[width=0.45\linewidth]{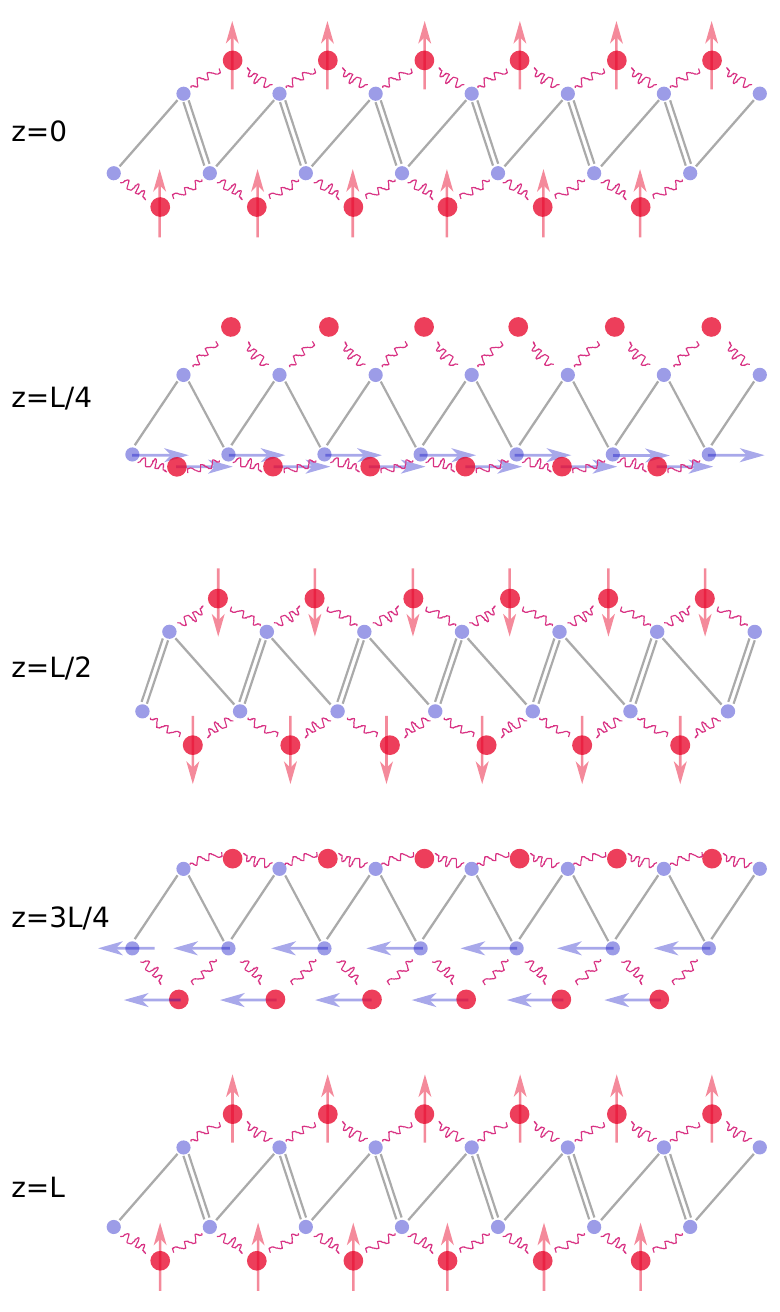}
    \caption{
    Geometry of the waveguide lattice and its modulations along the propagation direction. Detuned connector sites are highlighted by red, while arrows show the directions of local instantaneous shifts of the waveguides orthogonal to their axes. Solid lines and curves schematically show the nearest-neighbor and connector-site-mediated couplings between the waveguides. 
    }
    \label{FigS1}
\end{figure}

To realize the proposed tight-binding model for optical waveguides, we choose the particular geometry of the waveguides and their modulations summarized in Fig.~\ref{FigS1}. 
The correspondence between the tight-binding description of the waveguide lattice and the Schr{\"o}dinger equation ensures that the time $t = \tau T$ in the latter corresponds to the propagation distance $z = \tau L$ in the waveguide model, where $L$ is the spatial period of the modulation along the waveguides. 
Single (double) straight lines in  Fig.~\ref{FigS1} indicate the weaker (stronger) nearest neighbor couplings, while wavy lines indicate couplings to the red connector sites. 
The detuned connector sites create an effective coupling between the same-sublattice waveguides. For detunings appreciably larger then characteristic direct couplings, this physical mechanism is captured by the degenerate perturbation theory~\cite{Bir1974,Gorlach2018Mar}. 

The arrows in Fig.~\ref{FigS1} indicate the direction of instantaneous $z$-derivatives of the positions of the waveguides in the respective transverse planes, the absense of the arrow indicates zero derivative. Although specific modulations of waveguide coordinates along $z$ influences the exact dynamics and adiabaticity conditions, below we assume simple harmonic dependencies.

\begin{figure}[h!]
    \centering
    \includegraphics[width=0.2\linewidth]{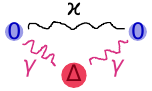}
    \caption{
    Trimer of waveguides connected by the detuned waveguide. The effective next-nearest-neighbor coupling between the two lattice waveguides (blue) mediated by the detuned site (red) is captured by the degenerate perturbation theory. On-site detunings are shown on top of the waveguides. 
    }
    \label{FigS2}
\end{figure}

In order to determine effective couplings and detunings from degenerate perturbation theory in the simplified model with couplings shown in Fig.~\ref{FigS1}, we consider a trimer of two lattice waveguides connected by a detuned waveguide, as shown in Fig.~\ref{FigS2}.

We analyze first the case of connector site with the detuned \textit{monopolar} $s$ mode. 
In this case, the trimer Hamiltonian reads 
\begin{equation}
\label{HamTrimer}
H_{3 \times 3}
=
\begin{pmatrix}
0 & \kappa & \gamma \\
\kappa & 0 & \gamma \\
\gamma & \gamma & \Delta 
\end{pmatrix}
. 
\end{equation}

We then transform this Hamiltonian to the basis of symmetric and antisymmetric modes $\{(-1,1,0)/\sqrt{2}, (1,1,0)/\sqrt{2}, (0,0,1)\}$,
\begin{equation}
    \tilde{H}_{3 \times 3}
    = 
    \left(
        \begin{array}{ccc}
         -\kappa  & 0 & 0 \\
         0 & \kappa  & \sqrt{2} \gamma  \\
         0 & \sqrt{2} \gamma  & \Delta  \\
        \end{array}
    \right). 
\end{equation}
$\tilde{H}_{3 \times 3}$ could be decomposed into the diagonal part $\tilde{H}_{3 \times 3}^{0} = \begin{pmatrix}
-\kappa & 0 & 0 \\
0 & \kappa & 0 \\
0 & 0 & \Delta 
\end{pmatrix}$ and the perturbation due to connector mode $\tilde{H}_{3 \times 3}' = \begin{pmatrix}
         0 & 0 & 0 \\
         0 & 0  & \sqrt{2} \gamma  \\
         0 & \sqrt{2} \gamma  & 0 
\end{pmatrix}$. 
Note that the symmetric connector mode couples only to the symmetric main-waveguide dimer mode. Similarly, anti-symmetric connector mode couples only to anti-symmetric main-waveguide dimer mode. 
Assuming $\Delta \gg \kappa, \gamma$, we obtain the $2 \times 2$ Hamiltonian $H_{2 \times 2}$ excluding the detuned $3^{\text{rd}}$ site by the degenerate perturbation theory~\cite{Bir1974,Gorlach2018Mar}: 
\begin{equation}
H_{2 \times 2 \,\, m m^{\prime}}
=
\tilde{H}_{3 \times 3 \,\, m m^{\prime}}
-
\frac{1}{2} \sum_s\left[\frac{1}{E_s^0-E_m^0}+\frac{1}{E_s^0-E_{m^{\prime}}^0}\right] \tilde{H}_{3 \times 3 \,\, m s}^{\prime} \tilde{H}_{3 \times 3 \,\, s m^{\prime}}^{\prime}
,
\end{equation}
where $m, m'$ indices are either $1$ or $2$, and $s=3$. 
Performing a unitary transformation back to the original basis of individual main waveguide modes, we recover the effective $2 \times 2$ Hamiltonian
\begin{equation}
\label{HamTrimer2}
H_{2 \times 2}
=
\begin{pmatrix}
\frac{\gamma^2}{\kappa - \Delta} & \kappa + \frac{\gamma^2}{\kappa - \Delta} \\
\kappa + \frac{\gamma^2}{\kappa - \Delta} & \frac{\gamma^2}{\kappa - \Delta} 
\end{pmatrix}
. 
\end{equation}
Importantly, this indicates not only next-nearest neighbor coupling $t = \kappa + \frac{\gamma^2}{\kappa - \Delta}$ mediated by the connector site, but also additional on-site energy shifts $u = \frac{\gamma^2}{\kappa - \Delta}$. These expressions for $u$ and $t$ suggest that in the proposed lattice the couplings $t_1, t_2$ and additional on-site energy shifts $u_1, u_2$ are modulated in-phase (synchronously). The corresponding Bloch Hamiltonian reads 
\begin{equation}
\label{Ham11}
H(k)=\begin{pmatrix}
2 t_2 \cos (k) + u_1 & J_1+J_2 e^{-i k} \\
J_1+J_2 e^{i k} & 2 t_1 \cos (k) + u_2
\end{pmatrix}
.
\end{equation}

\begin{figure}[h!]
    \centering
    \includegraphics[width=0.8\linewidth]{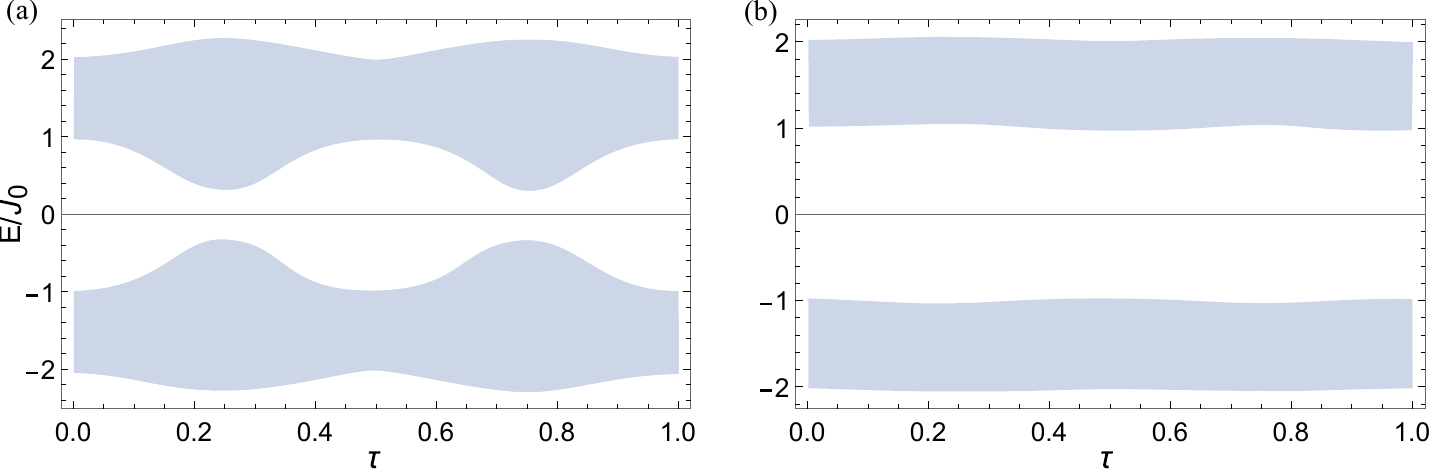}
    \caption{
    Modulated bulk spectra for symmetric connector modes, $s = 1$~(a), and antisymmetric connector modes, $s = -1$~(b). Other parameters are the same as in Fig.1 in the main text. Shaded regions mark the bulk bands. 
    }
    \label{modulated_spectra}
\end{figure}

Next, we consider another scenario of connector site with the detuned \textit{horizontal dipolar} $p_x$ mode. 
In this case, the trimer Hamiltonian reads 
\begin{equation}
\label{HamTrimerp}
H_{3 \times 3}
=
\begin{pmatrix}
0 & \kappa & \gamma \\
\kappa & 0 & -\gamma \\
\gamma & -\gamma & \Delta 
\end{pmatrix}
. 
\end{equation}
With similar assumptions $\Delta \gg \kappa, \gamma$, we obtain the following $2 \times 2$ Hamiltonian: 
\begin{equation}
\label{HamTrimerp2}
H_{2 \times 2}
=
\begin{pmatrix}
- \frac{\gamma^2}{\kappa + \Delta} & \kappa + \frac{\gamma^2}{\kappa + \Delta} \\
\kappa + \frac{\gamma^2}{\kappa + \Delta} & - \frac{\gamma^2}{\kappa + \Delta} 
\end{pmatrix}
. 
\end{equation}
We note that now NNN couplings $t_1, t_2$ and additional on-site energies $u_1, u_2$ are modulated {\it out-of-phase}. Hence, the two mechanisms of bandgap opening produced by $u_{1,2}$ and $t_{1,2}$ add up constructively, producing a larger bandgap and thus more robust transport. 
Indeed, consider a simple harmonic modulation as in the main text: 
\begin{align}
J_1(\tau) &= J_0 - \frac{A}{2}\cos(2\pi \tau)\:, \label{eq:J1}\\
J_2(\tau) &= J_0 + \frac{A}{2}\cos(2\pi \tau)\:, \label{eq:J2}\\
t_1(\tau) &= B \sin(2\pi \tau)\:, \label{eq:t1}\\
t_2(\tau) &= -B \sin(2\pi \tau)\:,\label{eq:t2}\\
u_1(\tau) &= s B \sin(2\pi \tau)\:, \label{eq:t1}\\
u_2(\tau) &= - s B \sin(2\pi \tau)\:,\label{eq:t2}
\end{align}
where, as follows from the analysis above, $s = +1$ for symmetric connector modes and $s = -1$ for antisymmetric connector modes. Note that for this example we choose modulations of $u_{1,2}$ with zero average since it does not affect the bandgap size~-- the central parameter setting the adiabaticity condition. As shown in the resulting modulated bulk spectra in Fig.~\ref{modulated_spectra} (with $J_0 = 1$ , $A = 1$, $B=0.35$, corresponding to a realistic situation where next-nearest-neighbor couplings are generally smaller than the nearest-neighbor ones), with all other parameters being equal, the case of antisymmetric connector modes has a substantially larger bandgap. We fully confirm this prediction below in tight-binding simulations using numerically extracted couplings, and also in full wave numerical simulations in Comsol Multiphysics. 

Below, we consider the two cases of symmetric-mode and antisymmetric-mode connector sites separately. For numerical simulations in COMSOL Multiphysics, we choose a typical scenario for optical waveguides fabricated in glass by the femtosecond laser-writing technique~\cite{Szameit2010Jul,Yan2024Oct,Wang2021Mar}: 
ambient glass refractive index $n = 1.48$ (borosilicate), elliptical waveguide profiles with a base contrast of $\delta n = 4 \cdot 10^{-4}$ plus additional contrast for connector sites specified in further sections, ellipse semi-axes $a=2.45\,\mu$m and $b=8.18\,\mu$m.


\section{II. Symmetric connector modes} 

\begin{figure}[h!]
    \centering
    \includegraphics[width=0.99\linewidth]{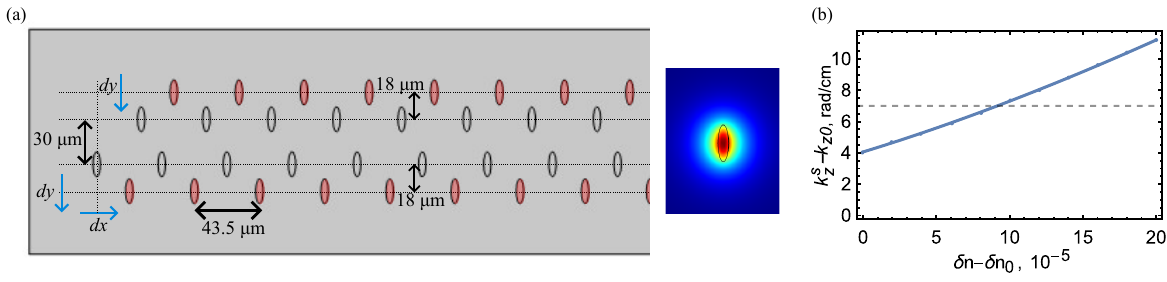}
    \caption{
    (a) Chosen geometry of the lattice and its spatial modulation amplitudes. The case of symmetric connector waveguide modes. Blue arrows show modulated parameters, while black double-arrows show fixed distances. 
    The coordinates of the lower main lattice waveguides and lower connector waveguides are modulated horizontally with amplitude $dx$, while both lower and upper connector waveguides are additionally modulated vertically with the amplitude $dy$. 
    Right panel: electric field amplitude profile of the mode in non-detuned lattice waveguides. 
    (b) Detuning $k_z^{s} - k_{z0} $ ($k_{z0} = 2\pi \cdot 1.48 / \lambda$ is the plane-wave wavenumber in bulk glass) of longitudinal wavenumber of $s$-mode in connector waveguides as a function of additional refractive index contrast $\delta n - \delta n_0$, where $\delta n_0 = 4 \cdot 10^{-4}$ is the contrast of main lattice waveguides. 
    The dashed line marks the target detuning of connector waveguide modes $\Delta = 3$~rad/cm with respect to the main waveguide modes. 
    }
    \label{FigS3}
\end{figure}

We choose a particular geometry of the lattice shown in Fig.~\ref{FigS3}(a). 
We choose simple harmonic modulations of the distances $dx$ and $dy$: 
\begin{eqnarray}
    \label{modulations}
    && dx = - 8 \cos{(2 \pi \tau)} \,\,\, \text{$\mu$m}, \\
    && dy = - 6 \sin{(2 \pi \tau)} \,\,\, \text{$\mu$m}. 
\end{eqnarray}
Furthermore, we choose detuning of the connector waveguides $\Delta = 3$~rad/cm by setting their refractive index contrast to $\delta n = 4.92 \cdot 10^{-4}$ [see Fig.~\ref{FigS3}(b)], roughly corresponding to the increase in the power of laser writing by $25 \%$ during the fabrication of connector waveguides.

We then extract the couplings $J$, $\kappa$ ($\gamma$) as functions of horizontal and vertical distances between the two main lattice waveguides (main lattice and connector waveguides) in corresponding dimers
from their eigenvalues $k_{z0} \pm \sigma$, where $k_{z0}$ is the propagation constant in isolated waveguides and $\sigma \in (J, \gamma, \kappa)$ is the corresponding coupling. 
Note that the coupling $\gamma$ is always calculated for perfectly degenerate main-lattice and connector-waveguide modes, as needed for its use in DPT theory to calculate $t$ and $u$ parameters. 

\begin{figure}[h!]
    \centering
    \includegraphics[width=0.98\linewidth]{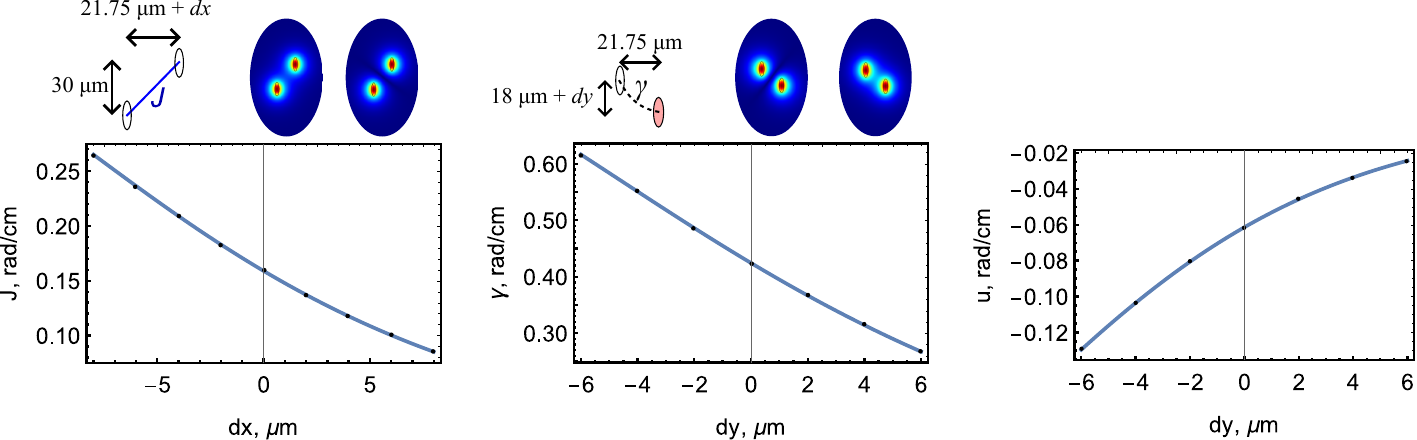}
    \caption{
    Couplings $J(dx)$, $\gamma(dy)$ and detuning $u(dy) = \gamma^2(dy)(\kappa - \Delta)$ dependencies on coordinate parameters $dx, dy$ extracted in COMSOL from corresponding waveguide dimer eigenvalues. 
    Insets show corresponding dimer geometries, as well as $|E|$ field profiles for symmetric and anti-symmetric dimer modes for the case $dx=dy=0$. 
    }
    \label{FigS3p5}
\end{figure}

Consistent with the lattice geometry and modulation scheme in Fig.~\ref{FigS3}, vertical distance for calculation of $J (dx)$ was fixed to $30 \,\mu\text{m}$ and horizontal distance was varied as $21.75 \,\mu\text{m} + dx$. 
Direct NNN coupling $\kappa = 0.052 \, \text{rad/cm}$ remains constant during the modulation at horizontal distance $43.5 \,\mu\text{m}$ and vertical distance $0$. 
For the calculation of main-connector coupling $\gamma (dy)$, the vertical distance was modulated as $18 \,\mu\text{m} + dy$ and horizontal distance was fixed at $21.75 \,\mu\text{m}$. 
The results are summarized in Fig.~\ref{FigS3p5}.

Then, the effective detunings $u_{1,2}$ and couplings $t_{1,2} = \kappa+u_{1,2}$ were constructed using the results of calculation outlined in the previous section in [Eq.~\eqref{HamTrimer2}]. 

\begin{figure}[h!]
    \centering
    \includegraphics[width=0.38\linewidth]{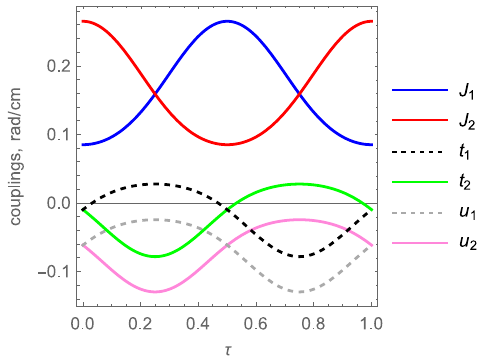}
    \caption{
    Modulation of the couplings and detunings for one modulation period. 
    }
    \label{FigS4}
\end{figure}

Finally, the couplings modulations as functions of $\tau = z/L$ are calculated by substituting specific modulations in Eq.~\eqref{modulations} and are shown in Fig.~\ref{FigS4}. 
Importantly, the obtained values of couplings justify the use of degenerate perturbation theory for a chosen detuning $\Delta$.

\begin{figure}[h!]
    \centering
    \includegraphics[width=0.65\linewidth]{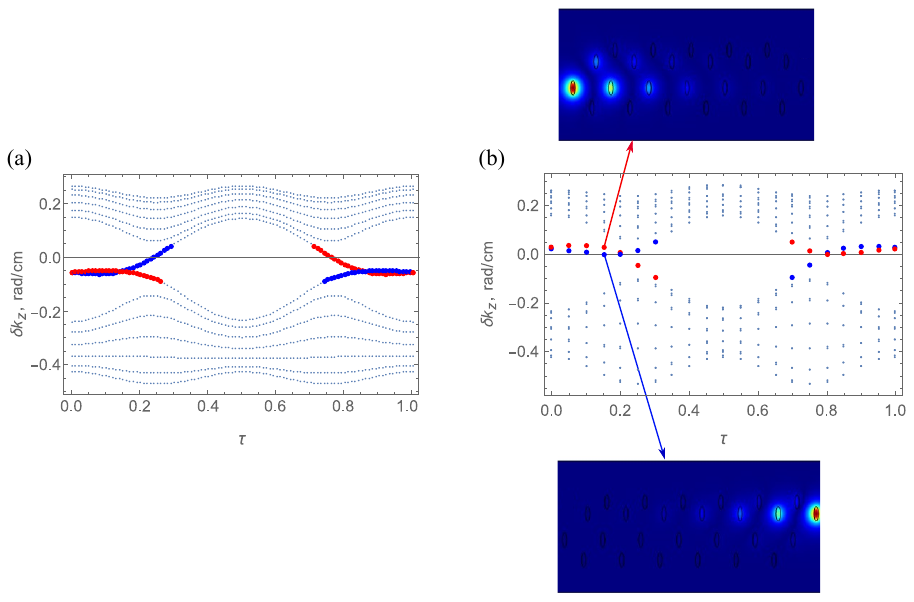}
    \caption{The spectrum of the propagation constants computed from the full tight-binding model with numerically retrieved couplings (a) and the results of full-wave simulations (b) for the finite lattice consisting of $7$ unit cells. Left- and right-localized edge states are highlighted by red and blue, respectively. 
    }
    \label{FigS5}
\end{figure}

Next, we calculate the spectra of the propagation constants in two ways. First, we examine the tight-binding model for a modulated finite lattice employing numerically computed couplings. Second, we evaluate the spectrum of the same lattice consisting of $7$ unit cells in Comsol Multiphysics, see Fig.~~\ref{FigS5}. Here, $\delta k_z = k_{z} - k_{z0}$, where $k_{z}$ is the collective mode longitudinal wavenumber. 

We find reasonably good correspondence between the numerical result and the tight-binding spectra, indicating applicability of the tight-binding description in our case. 

Importantly, both results show left- and right-localized topological edge states traversing the complete bandgap in the opposite directions, indicating nontrivial topological properties of the system.

\begin{figure}[h!]
    \centering
    \includegraphics[width=0.65\linewidth]{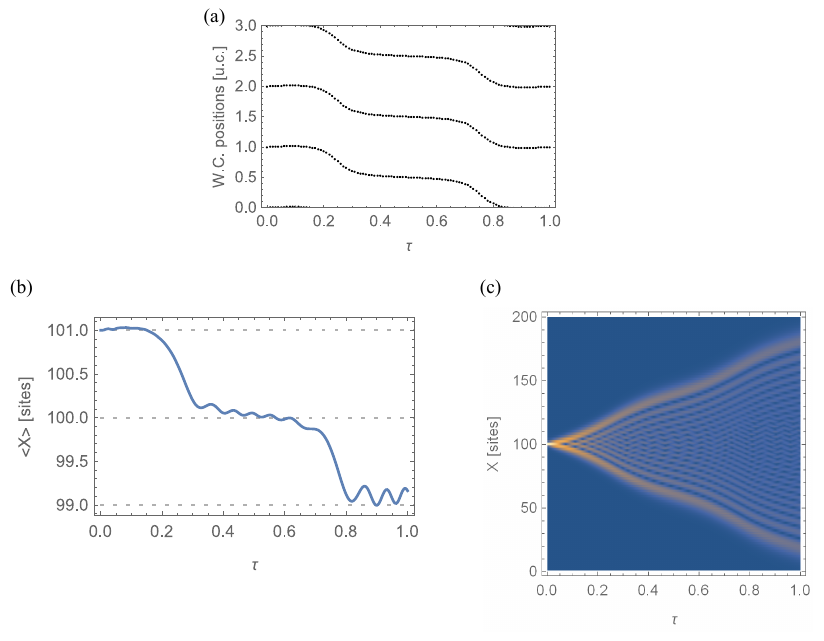}
    \caption{
    (a) Wannier center positions for the periodic Hamiltonian during one modulation cycle. 
    (b) Modulation of the coordinate of the center of mass for an initial point-like excitation constructed from the lower Bloch band during one period.   
    (c) Corresponding real-space discrete diffraction pattern. 
    }
    \label{FigS6}
\end{figure}

Using the tight-binding description of the proposed optical waveguide lattice, we also calculate the trajectories of the Wannier centers, see Fig.~\ref{FigS6}(a), which shows the shift by one unit cell during one pumping  period. The Thouless pumping simulation for the point-like initial excitation projected onto the lower Bloch band for the 
period $T = 300$~cm 
and array comprising $100$ unit cells 
fully supports this picture, see Fig.~\ref{FigS6}(b,c).

\clearpage
\newpage

\section{III. Anti-symmetric connector modes}

\begin{figure}[h!]
    \centering
    \includegraphics[width=0.99\linewidth]{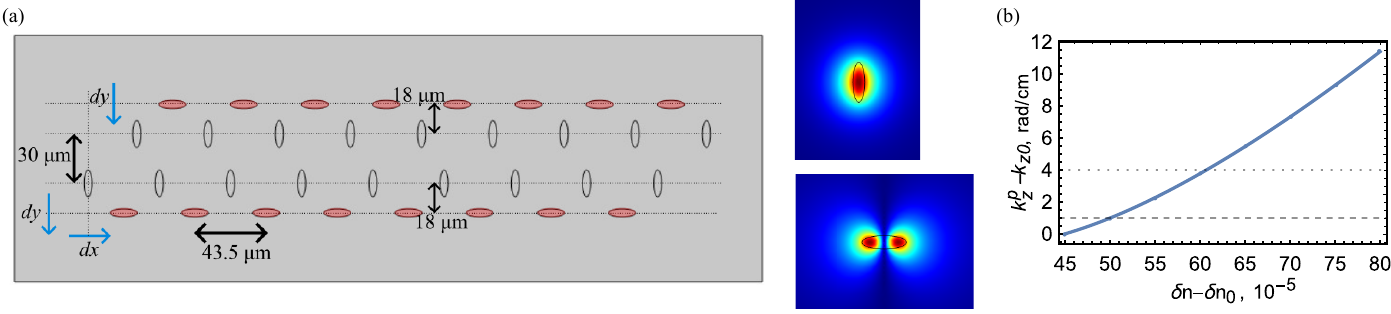}
    \caption{
    (a) Chosen geometry of the lattice and its spatial modulation amplitudes (anti-symmetric connector waveguide modes case). Right: electric field amplitude profile of the symmetric and antisymmetric modes in non-detuned vertical (upper panel) and detuned horizontal (lower panel) lattice waveguides, respectively. 
    (b) Detuning $k_z^{p} - k_{z0} $ of longitudinal wavenumber of $p$-mode in connector waveguides as a function of added refractive index contrast $\delta n - \delta n_0$, where $\delta n_0 = 4 \cdot 10^{-4}$ is the contrast of main lattice waveguides. 
    The dashed line marks the target detuning of connector waveguide modes $\Delta = - 3$~rad/cm with respect to main waveguide modes, 
    while dotted line~-- perfect degeneracy to the main lattice waveguides. 
    }
    \label{FigS3_p}
\end{figure}

Next we explore an alternative route when the connector sites host antisymmetric modes, which corresponds to the lattice geometry in Fig.~\ref{FigS3_p}(a). The  connector sites are now represented by the horizontally oriented waveguides supporting detuned $p_x$ modes. 
Importantly, such geometry facilitates minimal detrimental couplings between the connector waveguides (red) and the opposite-sublattice waveguides, considering the antisymmetric mode profile [see right panel of Fig.~\ref{FigS3_p}(a)], ensuring an excellent correspondence to the tight-binding model. 
We choose similar parameters, except for the new harmonic modulations of the $dy$ distances, $dy = - 4 \sin{(2 \pi \tau)} \,\,\, \text{$\mu$m}$. 
We choose detuning of the connector $p$-mode waveguides $\Delta = -3$~rad/cm by setting their refractive index contrast to $\delta n = 9.0 \cdot 10^{-4}$ [see Fig.~\ref{FigS3_p}(b)], which translates into the increase in the laser writing power by $125 \%$ during the fabrication of connector waveguides with respect to the main lattice waveguides. Note that perfect degeneracy between $s$ in main waveguides and $p_x$ modes in connector waveguides corresponds to contrast in the latter of $\delta n = 10.07 \cdot 10^{-4}$.

\begin{figure}[h!]
    \centering
    \includegraphics[width=0.70\linewidth]{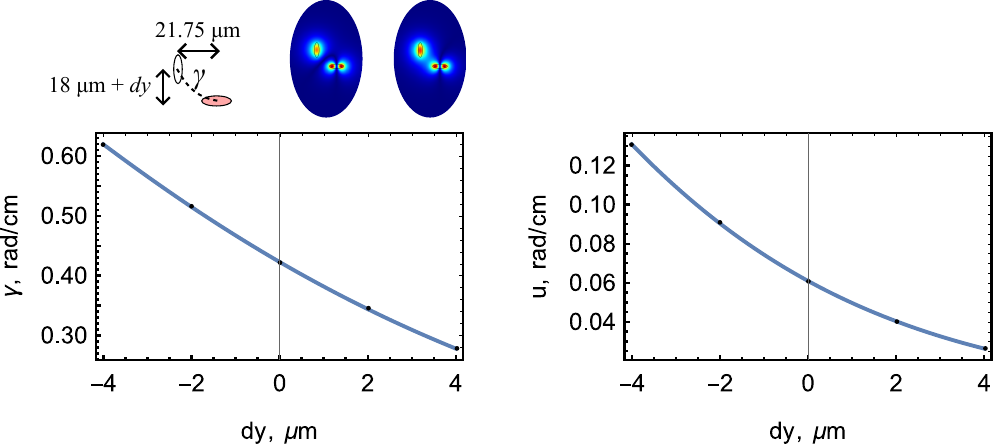}
    \caption{
    Coupling $\gamma(dy)$ and detuning $u(dy) = - \gamma^2(dy)(\kappa + \Delta)$ dependencies on coordinate parameters $dx, dy$ extracted in COMSOL from corresponding waveguide dimer eigenvalues. $J(dx)$ is the same as in Fig.~\ref{FigS3p5}. 
    Insets show corresponding dimer geometry, as well as $|E|$ field profiles for symmetric and anti-symmetric dimer modes for the case $dx=dy=0$. 
    }
    \label{FigS3p5_p}
\end{figure}

Due to similarity of geometry with the case with symmetric connector modes (cf. Fig.~\ref{FigS3}), the model with anti-symmetric connector modes features the same $J (dx)$ and direct NNN coupling $\kappa$. 
%
%
For the main-connector coupling $\gamma (dy)$, the vertical distance is modulated as $18 \,\mu\text{m} + dy$ and horizontal distance is fixed at $21.75 \,\mu\text{m}$. 
The effective detunings $u_{1,2}$ and couplings $t_{1,2} = \kappa - u_{1,2}$ are calculated according to Eq.~\eqref{HamTrimerp2}. 
The couplings and detunings modulations as functions of $\tau = z/L$ are calculated by substituting specific modulations $dx = - 8 \cos{(2 \pi \tau)} \,\,\, \text{$\mu$m}$, $dy = - 4 \sin{(2 \pi \tau)} \,\,\, \text{$\mu$m}$ and depicted in Fig.~\ref{FigS3p5_p}.


\begin{figure}[h!]
    \centering
    \includegraphics[width=0.38\linewidth]{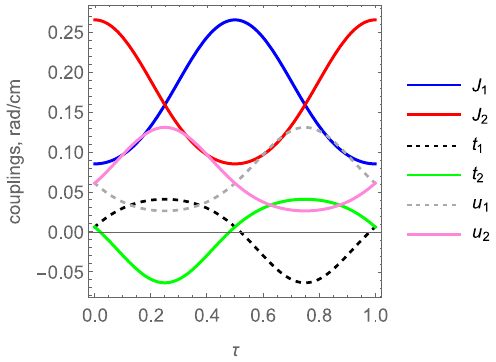}
    \caption{
    Modulation of the couplings for one period extracted from COMSOL dimer eigenvalues. 
    }
    \label{FigS4_p}
\end{figure}

\begin{figure}[h!]
    \centering
    \includegraphics[width=0.65\linewidth]{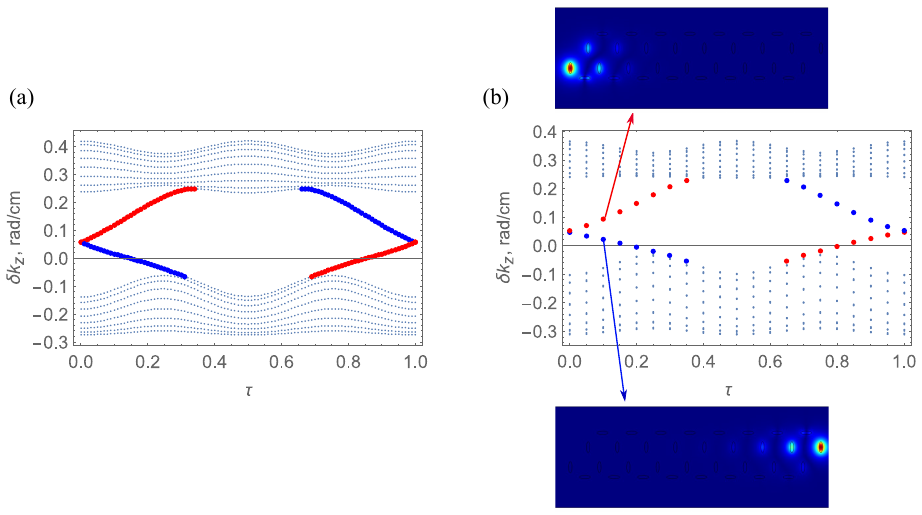}
    \caption{
    The spectrum of the propagation constants calculated from the full tight-binding model with numerically retrieved couplings (a) and the results of full-wave simulations in Comsol Multiphysics (b) for the finite lattice consisting of $9$ unit cells. Left- and right-localized edge states are highlighted by red and blue, respectively. 
    }
    \label{FigS5_p}
\end{figure}

The modulated finite-lattice tight-binding with numerically calculated couplings and fully numerical spectrum for a lattice comprising $9$ unit cells is shown in Fig.~\ref{FigS5_p}. 
We find exceptional correspondence between the numerical result and the tight-binding spectra. 
The spectrum in this case features a particularly large bandgap, facilitating quantized transport for smaller lattices and shorter periods with greater disorder resilience. 
This is directly connected to the fact that in this case of anti-symmetric connector modes the two mechanisms of bandgap opening produced by $u_{1,2}$ and $t_{1,2}$ add up. 
%


\begin{figure}[h!]
    \centering
    \includegraphics[width=0.85\linewidth]{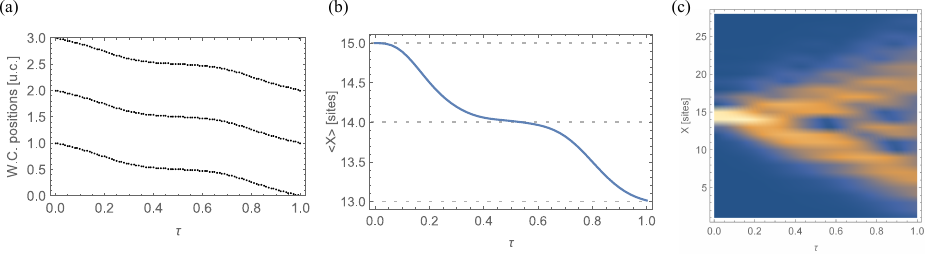}
    \caption{
    (a) Wannier center positions for the periodic Hamiltonian during one modulation cycle. 
    (b) Modulation of the coordinate of the center of mass for an initial point-like excitation constructed from the lower Bloch band during one period.   
    (c) Corresponding real-space discrete diffraction pattern. 
    }
    \label{FigS6_p}
\end{figure}


The path of Wannier centers shown in Fig.~\ref{FigS6_p}(a) reveals nontrivial topology with a remarkably smoother trajectory of Wannier centers. 
The Thouless pumping simulation for the point-like initial excitation projected onto the lower Bloch band for the 
smaller period $T = 50$~cm 
and array comprising only $14$ unit cells 
clearly shows quantized transport for the center of mass of the intensity distribution, see Fig.~\ref{FigS6_p}(b,c).

\begin{figure}[h!]
    \centering
    \includegraphics[width=0.80\linewidth]{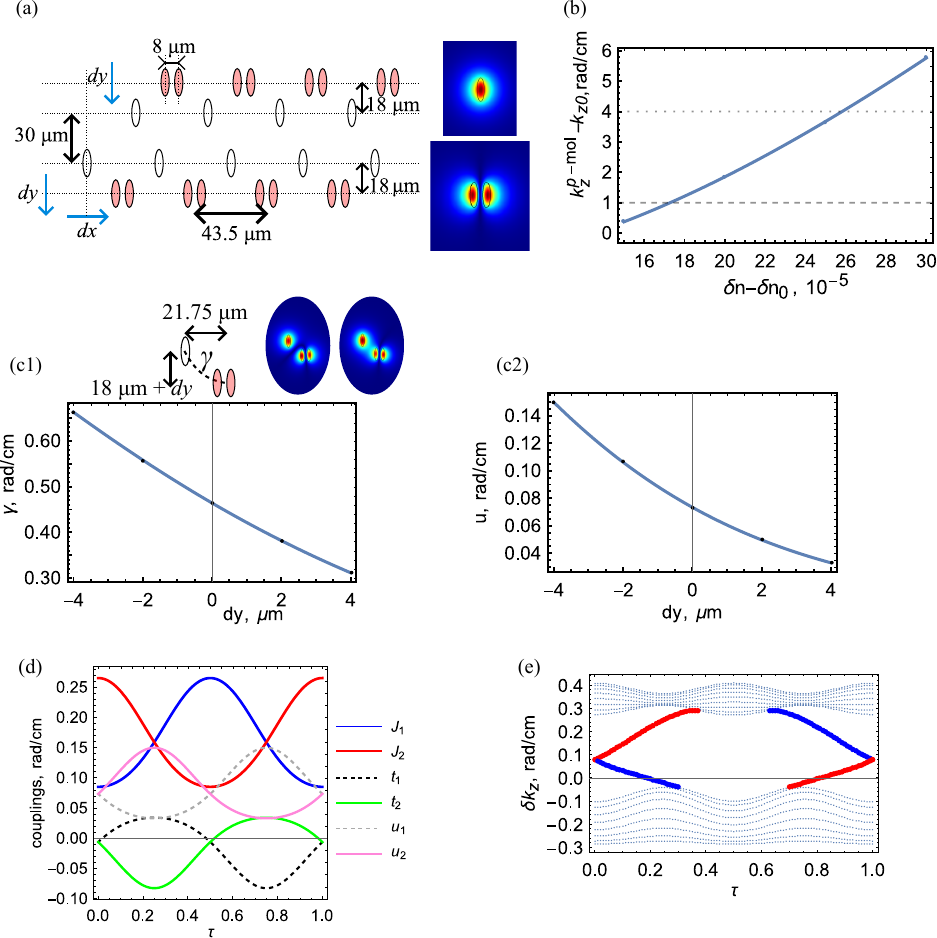}
    \caption{
    Results for the lattice with antisymmetric photonic molecule connector modes. 
    (a-b) Analogues of Fig.~\ref{FigS3_p}(a-b). 
    (c1-c2) Analogues of Fig.~\ref{FigS3p5_p}. 
    (d) Analogue of Fig.~\ref{FigS4_p}. 
    (e) Analogue of Fig.~\ref{FigS5_p}(a). 
    }
    \label{FigS3_p-mol}
\end{figure}

An important alternative to using horizontally oriented waveguides with $p$-modes is to fabricate corresponding fine-tuned photonic molecules comprising pairs of closely placed vertically oriented waveguides~\cite{Mazanov2024Apr}. 
For this design, we choose photonic molecules made from pairs of geometrically identical waveguides placed at horizontal distance~$8\,\mu$m, see Fig.~\ref{FigS3_p-mol}, and tune their refractive index contrast for near-degeneracy to the $s$-modes, which appears at $\delta n=6.58 \cdot 10^{-4}$, while detuning $\Delta = -3$ is achieved for $\delta n=5.73 \cdot 10^{-4}$, corresponding  to the relative increase in writing power by $40 \%$ with respect to the main lattice waveguides. 
In all other respects, we keep the same lattice geometry and modulations amplitudes. 
Interestingly, we obtain nearly identical results for this case in terms of effective couplings and finite lattice spectrum as in the case of horizontal connector waveguides. 
Thus, such photonic molecule approach can be applied on the same footing for the same lattice geometry and modulations, so that its important advantage in conventional writing from one wafer facet could be fully harnessed.

\clearpage
\newpage

\subsection{Tight-binding model with connector modes included explicitly} 

Here we explore the pumping protocol in the tight-binding model which includes the connector $p_x$-modes explicitly, in order to validate the applicability of the degenerate perturbation theory. 

For convenience, we use the basis of sites in the following order for the lattice of $N$ unit cells: $2N$ main lattice site modes followed by $2(N-1)$ modes of the connector-site modes. Note that in order to obtain a smooth center-of-mass trajectory, the necessary projector $\hat{P}$ for the initial state is constructed using the eigenstates below the bulk bandgap but above the bands corresponding to the connector sites.

\begin{figure}[h!]
    \centering
    \includegraphics[width=0.70\linewidth]{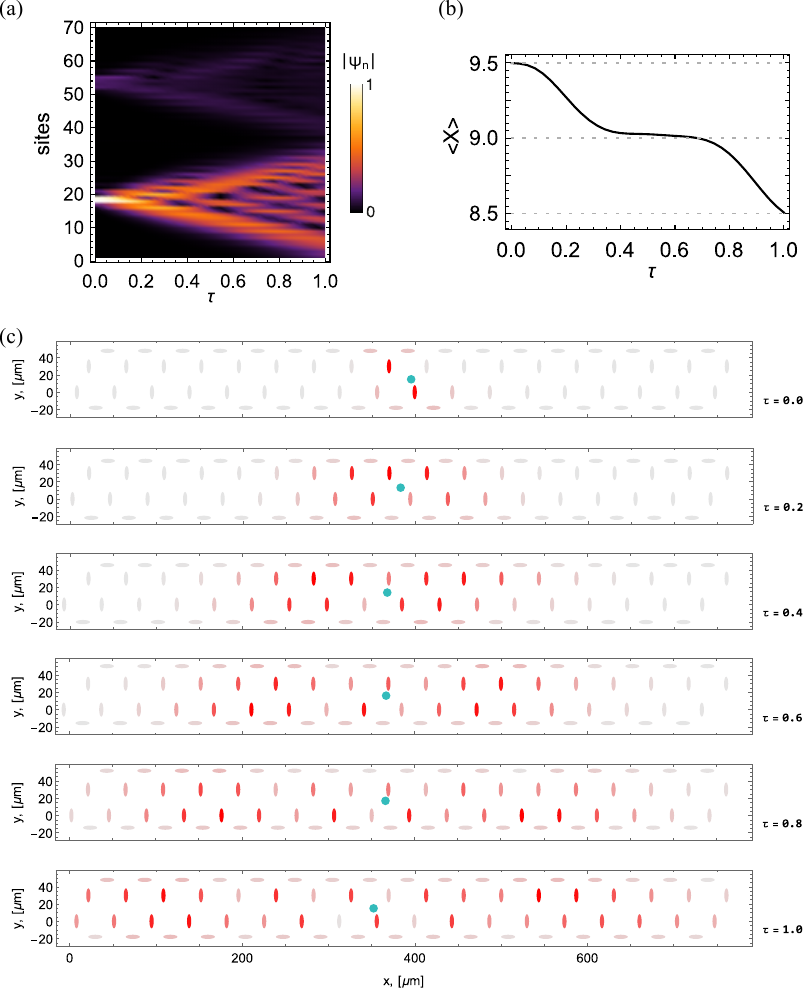}
    \caption{
    Results for the tight-binding model including connector site modes explicitly. 
    (a) Distribution of the absolute value of the amplitude during pumping. 
    (b) Shift of center-of-mass of the initial wavepacket. 
    (c) Corresponding snapshots of the absolute values of the amplitude during the pumping process ($\tau$ is indicated on the right to the plots). 
    The blue dot shows the center of mass defined as a sum where each mode amplitude is weighted by the geometric center-of-mass coordinate of its corresponding unit cell. 
    Parameters: $18$ unit cells, $T = 35$~cm; other parameters are the same as in Fig.4 in the main text. 
    }
    \label{model2+snaps}
\end{figure}

Fig.~\ref{model2+snaps}(a,b) shows that amplitudes at the connector site modes during pumping are more that $5$ times smaller compared to ones in the main lattice sites (appearing at the top half of the distribution in Fig.~\ref{model2+snaps}(a)), validating the use of simplified DPR theory with only main-lattice sites, while the accuracy of the center-of-mass shift quantization is similar to Fig.4 in the main text.

\clearpage
\newpage

\section{IV. Implementation of the pumping scheme for polariton condensates}

Besides optical implementation of our model, we briefly discuss another design applicable to the coupled lattices of polariton condensates.

Optical pumping generates clouds of electrons and holes that form an incoherent excitonic reservoir in quantum wells embedded in a microcavity. The role of this reservoir is two-fold: (1) it creates a repulsive potential for exciton-polaritons with a magnitude proportional to the exciton density, (2) it feeds the population of the polariton mode due to the inelastic exciton scattering. To demonstrate the quantized transport of a bosonic condensate of exciton-polaritons one would need to complement the reservoir pumping by a resonant optical pumping of the polariton mode. A polariton wave-packet created by the resonant laser pulse would be stabilised in time and preserved from decay by a stimulated scattering of exciton-polaritons from the exciton reservoir. In turn, the dynamics of the polariton condensate generated this way may be observed experimentally by means of time- and space-resolved photoluminescence spectroscopy, see e.g. Ref.~\cite{Amo2009}.

The Thouless pumping scheme and its generalizations for polariton condensates can be implemented  by interfering the two pump beams with the slightly detuned in-plane momenta $k$ and $k+\Delta k$ as well as detuned frequencies $\om$ and $\om+\Delta\om$ which create a moving potential for polaritons. Below, we elaborate the conceptual scheme of such an experiment.

\begin{figure}[h!]
    \centering
    \includegraphics[width=0.60\linewidth]{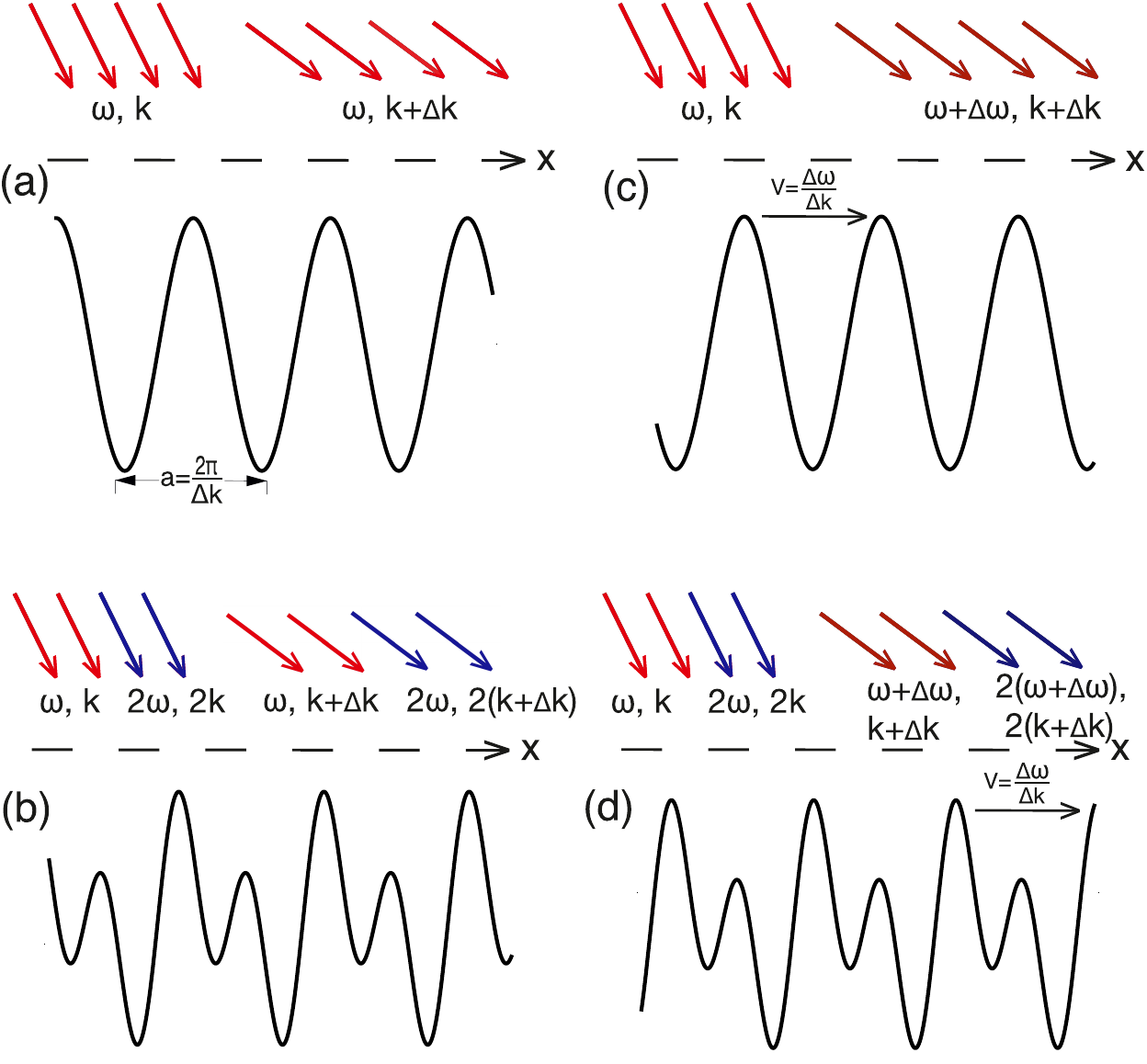}
    \caption{Conceptual implementation of the Thouless-type pumping for polariton condensate. (a) Sinusoidal static lattice obtained by interfering the two waves with the same frequency and different $x$-projections of the wave vector. (b) Su-Schrieffer-Heeger-type lattice obtained by interfering the waves with frequencies $\omega$ and $2\om$ and wave vector projections $k$, $k+\Delta k$, $2\,k$, $2\,(k+\Delta k)$. (c) Moving sinusoidal lattice obtained by using the waves with the frequency detuning $\Delta\om$. (d) Moving Su-Schrieffer-Heeger-type lattice realizing the Thouless pumping scheme.
    }
    \label{FigS12}
\end{figure}

(a) If the two pump beams have identical frequencies and amplitudes, but slightly detuned propagation constants, the resulting field of a pump takes the form
\begin{equation}
    E(x)=A\,e^{ikx-i\om t}+A\,e^{i(k+\Delta k)x-i\om t}\:,
\end{equation}
where an inessential relative phase between the two beams can be removed by the suitable choice of the coordinate origin. The respective intensity distribution then reads:
\begin{equation}
    I(x)\propto |E(x)|^2=2A^2\,\left[1+\cos\left(\Delta k x\right)\right]\:,
\end{equation}
which provides a sinusoidal effective potential for polaritons, Fig.~\ref{FigS12}(a).

(b) A Su-Schrieffer-Heeger type of lattice can be implemented using the pump beams with the two frequencies $\om$ and $2\om$ such that the total electric field of the pump takes the form
\begin{equation}
    E(x)=A_1\,e^{ikx-i\om t}+A_1\,e^{i(k+\Delta k)x-i\om t}+A_2\,e^{2ikx-2i\om t}+A_2\,e^{2i(k+\Delta k)x-2i\om t+i\phi}\:.
\end{equation}
Such field corresponds to the intensity distribution of the form
\begin{equation}
    I(x)\propto \left<|E(x)|^2\right>=2A_1^2\,\left[1+\cos\left(\Delta k x\right)\right]+2A_2^2\,\left[1+\cos\left(2\Delta k x+\phi\right)\right]\:,
\end{equation}
where after calculating $|E(x)|^2$ we drop rapidly oscillating terms. Now the shape of the effective potential is controlled by the two independent parameters: the ratio of $A_1$ and $A_2$ amplitudes as well as the relative phase $\phi$. This allows to create a lattice resembling the Su-Schrieffer-Heeger model with the tunable dimerization strength~\cite{Atala2013} as illustrated in Fig.~\ref{FigS12}(b).

(c) To implement a moving potential, the frequencies of the two pump beams need to be slightly detuned. Specifically, we consider a pump of the form
\begin{equation}
    E(x)=A\,e^{ikx-i\om t}+A\,e^{i(k+\Delta k)x-i(\om+\Delta\om)t}\:.
\end{equation}
After averaging over time which excludes rapidly oscillating contributions, we recover the intensity distribution
\begin{equation}    I(x)\propto\left<|E(x)|^2\right>=2A^2\,\left[1+\cos\left(\Delta k x-\Delta \om t\right)\right]\:.
\end{equation}
This corresponds to the sinusoidal potential slowly moving in space with the controllable speed $v=\Delta\om/\Delta k$, Fig.~\ref{FigS12}(c). Recently, this approach has been employed to engineer nonreciprocal band structures of exciton-polaritons~\cite{Fraser2024}.

(d) Finally, the two ideas~-- SSH-type lattice and slowly drifting potential can be combined together by introducing a pump
\begin{equation}
E(x)=A_1\,e^{ikx-i\om t}+A_1\,e^{i(k+\Delta k)x-i(\om+\Delta\om) t}+A_2\,e^{2ikx-2i\om t}+A_2\,e^{2i(k+\Delta k)x-2i(\om+\Delta\om) t+i\phi}\:,
\end{equation}
which results in the intensity distribution of the form
\begin{equation}
I(x)\propto \left<|E(x)|^2\right>=2A_1^2\,\left[1+\cos\left(\Delta k x-\Delta\om t\right)\right]+2A_2^2\,\left[1+\cos\left(2\Delta k x-2\Delta\om t+\phi\right)\right]\:.    
\end{equation}
This creates a dimerized SSH-type lattice in space at each moment of time, and this lattice moves with the speed $v=\Delta\om/\Delta k$, Fig.~\ref{FigS12}(d). In turn, such moving potential realizes an instance of the Thouless-like pump.

\section{V. Connection between the Wannier center displacement and the Chern number}


In this section, we consider an adiabatic Thouless pump assuming the periodic variation of the Hamiltonian $\hat{H}(k, t)$ in time. Following Ref.~\cite{Asboth2016}, we revisit the connection between the displacement of the wavepacket center of mass and the space-time Chern number for the bands of the instantaneous spectrum.

The expectation value of the position for the wavepacket formed by the eigenstates from the $n$-th band is given by
\begin{equation}
x(t) = \frac{i}{2\pi} \int_{-\pi}^{\pi} \langle u_n(k,t) | \partial_k u_n(k,t) \rangle  dk,
\end{equation}
where $u_n(k,t)$ are the Bloch functions of the $n^{\text{th}}$ band.

The total displacement of the Wannier center over one pumping cycle reads
\begin{equation}
\Delta x_{0,T} = x(T) - x(0) = \frac{i}{2\pi} \left[ \int \langle u_n(k,T) | \partial_k u_n(k,T) \rangle dk - \int \langle u_n(k,0) | \partial_k u_n(k,0) \rangle dk \right].
\end{equation}

To analyze this expression, we discretize the time interval $[0,T]$ into $N$ small segments of duration $\Delta t = T/N$, with $t_m = m\Delta t$. The total displacement can then be expressed as a sum of infinitesimal displacements:
\begin{equation}
\Delta x_{0,T} = \lim_{N \rightarrow \infty} \sum_{m=0}^{N-1} \Delta x_{t_m, t_m+\Delta t},
\end{equation}
where the infinitesimal displacement for a single time step is:
\begin{equation}
\Delta x(t_m,t_m+\Delta t) = \frac{i}{2 \pi}\int_{-\pi}^{\pi} dk \left[ \langle u_n(t_m+\Delta t) | \partial_k u_n(t_m+\Delta t) \rangle - \langle u_n(t_m) | \partial_k u_n(t_m) \rangle \right].
\end{equation}

We can rewrite this expression using the Berry connection:
\begin{equation}
    A_k^{(n)}(k,t) = i \langle u_n(k,t) | \partial_k u_n(k,t) \rangle\:.
\end{equation}
Therefore, the displacement formula simplifies to:
\begin{equation}
\Delta x(t_m, t_m + \Delta t) = \frac{1}{2\pi} \left[ \int A^{(n)}_k(k,t_m + \Delta t)  dk - \int A^{(n)}_k(k,t_m)  dk \right].
\end{equation}

\begin{figure} [h!]
    \centering
    \includegraphics[width=0.4\linewidth]{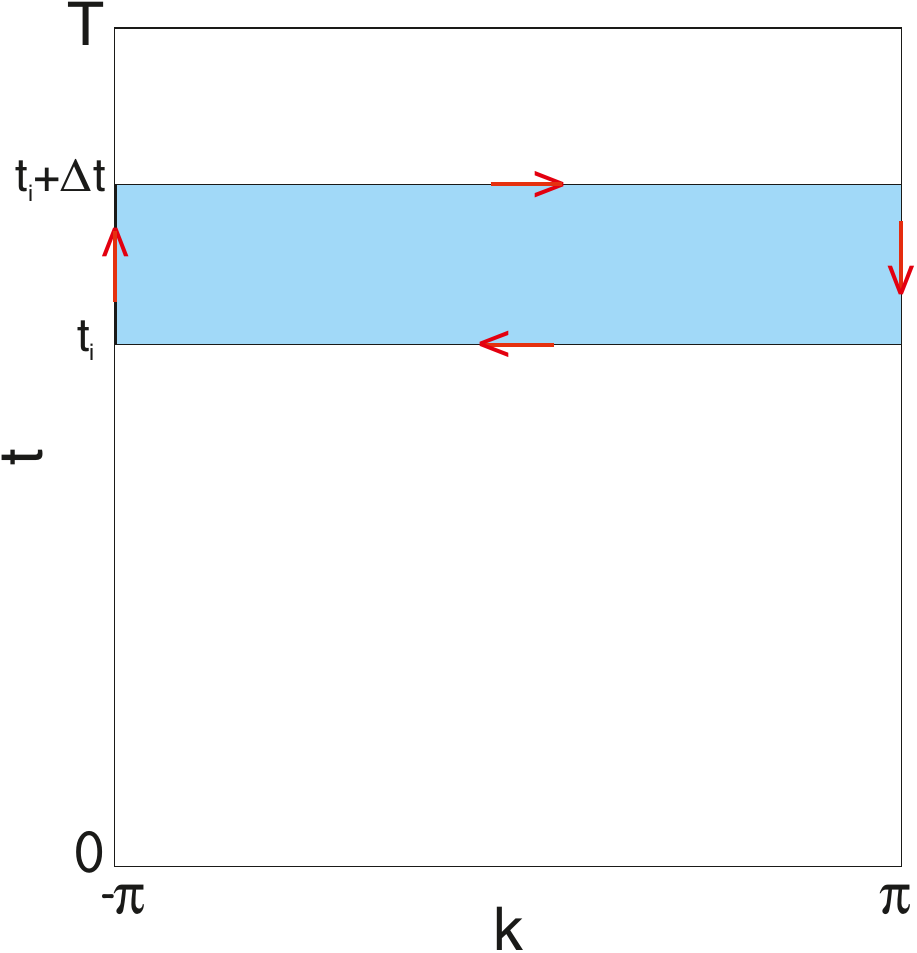}
    \caption{Schematic showing the calculation of the Chern number in $k$-$t$ space. Time variable $t$ plays the role of the synthetic momentum.}
    \label{FigS17}
\end{figure}

Next we consider a small rectangle $R_m = [-\pi, \pi) \times [t_m, t_m + \Delta t]$ in the $(k,t)$ space, as shown in Fig.~\ref{FigS17}.

The contour integral of the full Berry connection $\mathbf{A}^{(n)} = (A^{(n)}_k, A^{(n)}_t)$ around its boundary $\partial R_m$ is:
\begin{align}
\oint_{\partial R_m} \mathbf{A}^{(n)} \cdot d\mathbf{R} &= \int_{t_m}^{t_m+\Delta t} A^{(n)}_t(-\pi,t) dt + \int_{-\pi}^{\pi} A^{(n)}_k(k,t_m+\Delta t) dk \\
&\quad + \int_{t_m+\Delta t}^{t_m} A^{(n)}_t(\pi,t) dt + \int_{\pi}^{-\pi} A^{(n)}_k(k,t_m) dk.
\end{align}
Due to the periodicity of the Brillouin zone,  $A^{(n)}_t(-\pi,t) = A^{(n)}_t(\pi,t)$ and the time integrals cancel out. Hence, we recover:
\begin{equation}
\oint_{\partial R_m} \mathbf{A}^{(n)} \cdot d\mathbf{R} =\int_{-\pi}^{\pi} A^{(n)}_k(k,t_m+\Delta t)\, dk - \int_{-\pi}^{\pi} A^{(n)}_k(k,t_m)\, dk.
\end{equation}
Comparing this to the previous expression for the displacement, we derive:
\begin{equation}
\Delta x(t_m, t_m + \Delta t) = \frac{1}{2\pi} \oint_{\partial R_m} \mathbf{A}^{(n)} \cdot d\mathbf{R}.
\end{equation}
Next we define the Berry curvature by the expression
\begin{equation}
    B^{(n)} = \partial_k A^{(n)}_t - \partial_t A^{(n)}_k
\end{equation}
and, applying Stokes' theorem, we recover
\begin{equation}
\oint_{\partial R_m} \mathbf{A}^{(n)} \cdot d\mathbf{R} = \int_{\partial R_m} B^{(n)}(k,t)\,  dk\,  dt.
\end{equation}
Hence, the infinitesimal displacement becomes:
\begin{equation}
\Delta x(t_m, t_m + \Delta t) = \frac{1}{2\pi} \int_{\partial R_m} B^{(n)}(k,t) dk dt,
\end{equation}
while the total displacement over the cycle is:
\begin{equation}
\Delta x_{0,T} = \lim_{N\to\infty} \sum_{m=0}^{N-1} \Delta x(t_m, t_m + \Delta t) = \frac{1}{2\pi} \int\int_{BZ}  B^{(n)}(k,t)dk dt.
\end{equation}
The double integral on the right-hand side is precisely the Chern number $C_n$ for the $n$-th band over the closed $(k,t)$ manifold:
\begin{equation}
C_n = \frac{1}{2\pi}\int\int_{BZ}  B^{(n)}(k,t)dk dt.
\end{equation}
Therefore:
\begin{equation}
\Delta x_{0,T} = C_n.
\end{equation}

Thus, the displacement of the Wannier centers per driving cycle is equal to the Chern number defined in the two-dimensional space of wave number and time. In our case, the displacement of the Wannier centers by one lattice period during one pumping cycle implies Chern number equal to 1.

\bibliography{QuantizedTransportSupplement.bib}